\documentstyle[12pt,twoside,fleqn,espcrc1,epsfig,glas]{article}

\newlength{\dinwidth}
\newlength{\dinmargin}
\newcommand{\AmS}{{\protect\the\textfont2
  A\kern-.1667em\lower.5ex\hbox{M}\kern-.125emS}}

\newcommand{\sgeq}{\raisebox{-0.5mm}{$\stackrel{>}{\scriptstyle{\sim}}$}}

\newcommand{\an}{\overline{\alpha}_0}
\newcommand{\as}{\alpha_s}
\newcommand{\asmz}{\alpha_s(M_Z)}

\newcommand{\ee}{\mbox{$e^+e^-$}}
\newcommand{\mean}[1]{\left< #1 \right>}
\newcommand{\fmean}{\mean{F}}
\newcommand{\fpert}{\fmean^{\rm pert}}
\newcommand{\fpow}{\fmean^{\rm pow}}

\newcommand{\mi}{\mu_{\scriptscriptstyle I}}

\newcommand{\anmi}{\overline{\alpha}_0(\mi)}

\newcommand{\rbthm}{\rule[-2ex]{0ex}{5ex}}

\newcommand{\rbtnpbr}{\rule[-1.2ex]{0ex}{3.6ex}}

\newcommand{\eprint}[2]{{hep-#1/}#2.}

\begin{document}

\begin{titlepage}{GLAS-PPE/99--14}{25$^{\underline{\rm{th}}}$ November 1999}
\title{Event Shapes and Forward Jet Production at HERA}

\author{A.T. Doyle
%\address{Dept. of Physics and Astronomy, 
%        University of Glasgow, \\ 
%        Glasgow G12 8QQ, United Kingdom}
}
       
%\maketitle

\begin{abstract}
Analyses of event shapes and forward jet production in deep inelastic scattering at the HERA collider are described. The
results are compared to QCD predictions. 

\vspace{1.5cm}
\centerline{\em Talk presented at the Hadron99 Conference,}
\centerline{\em Beijing, August 1999.}
\end{abstract}
\end{titlepage}

%\section{INTRODUCTION: QCD PUZZLES}
\section{EVENT SHAPES}
A recent revival of interest in the study of event shape measurements has 
been prompted by theoretical developments in the understanding of hadronisation
or power corrections. Here, perturbative QCD calculations are extended into 
the region of low momentum transfers using approximations to higher-order 
graphs.
At HERA, the $Q^2$ scale can be varied over four orders of magnitude enabling
power corrections (proportional to $1/Q^p$, where $p$ is the power) to be 
studied in detail. 
The data are compared to theoretical expectations for the power 
corrections, characterised by 
an effective coupling $\an(\mi)$ specified at the infra-red
matching scale $\mi \simeq 2$~GeV,
and NLO pQCD calculations, determined by $\as(M_Z)$.

The Breit frame is where the exchanged gauge boson in DIS is 
purely space-like. Viewed in this frame, the incoming QPM quark 
is back-scattered with equal and opposite momentum ($Q/2$).
Event shapes have been measured in the direction of the struck quark, 
corresponding to the current region, which is directly
analogous to one hemisphere of a purely time-like \ee~interaction.  
A series of event shape variables have been studied which have varying 
sensitivity to hadronisation/power corrections and are noted below. 
Thrust ($\tau= 1 - T_{z}$)
and jet broadening ($B$) sum the longitudinal and transverse momenta, 
respectively, of individual particles along the photon axis. 
$\tau_c$ is also defined with respect to the reconstructed thrust axis,
in order to compare directly with \ee.
Jet mass ($\rho$) and the $C$ parameter sum the product of two particle momenta 
weighted by $1-\cos\theta_{ij}$ and $\sin^2\theta_{ij}$, respectively.
%Differential two-jet rates, $y_{fJ}$ and $y_{k_t}$ denote the transition values 
%of an event from $(2+1)\rightarrow(1+1)$ jets for the JADE and $k_t$ algorithms.
%These variables each characterise a given event configuration 
%in the current region of the Breit frame. 
These variables are each measured 
in the current region of the Breit frame
and are scaled by the total visible energy ($E_{vis}$).

Mean values of these event shapes haves been measured by H1~\cite{H1event}
and ZEUS~\cite{ZEUSevent} which can be interpreted 
theoretically via
$$\fmean =\fpert + \fpow$$
where
$$  \fpert = c_{1,F}(x,Q)\cdot\as(Q) + c_{2,F}(x,Q)\cdot\as^2(Q)$$
with $c_{1,F}(x,Q)$ and $c_{2,F}(x,Q)$ being calculable coefficients of pQCD
determined by the DISENT NLO program~\cite{DISENT} 
with renormalisation/factorisation scales set at $\mu_R=\mu_F=Q$.
The power corrections are 
proportional to $1/Q$
%, for all cases except $\mean{y_{k_t}}$, 
with 
$$  \fpow = {\cal M}^\prime
  \frac{a_F}{Q} \frac{16\mi}{3\pi} \bigg[ \an(\mi) - \as(Q)
  - \frac{23}{6\pi}\left( \ln\frac{Q}{\mi} + 1.45\right) \as^2(Q) \bigg].$$
Here $a_F$ is a calculable $F$-dependent coefficient and 
${\cal M}^\prime = 2{\cal M}/\pi \simeq 1.14$~\cite{Milan1}
%${\cal M} \simeq 1.79$~\cite{Milan1}
\footnote{The latest theoretical evaluation determines ${\cal M}^\prime 
\simeq 0.95$~\cite{Milan2}.}
 is a two-loop level refinement to the calculations
called the Milan factor.

The results of two-parameter fits to the H1 data are given in Table~1 and 
Figure~1(left). The $\chi^2$ per degree of freedom are consistent with unity.
The correlation of $\anmi$ and $\asmz$ is high and negative for 
$\mean{\tau}$ but is less for other variables.
Results from similar fits (excluding the Milan factor) to ZEUS charged-hadron 
data and the earlier H1 published data are shown in Figure~1(right). 
In each case, the data for the means of $\tau$, $B$, $\tau_C$, 
$C$ and $\rho$,
\footnote{
Note that the H1 published result for jet mass is 
normalised to $Q^2$ rather than $E_{vis}^2$.}
are described reasonably well by the theory, with a 
value of the universal non-perturbative parameter $\anmi$ of 
$\simeq 0.5 \pm 20\%$ and a value of $\asmz$ close to the world average.
%The latest H1 preliminary result tends towards higher values of $\asmz$. 
In conclusion, the theoretical 
approach works remarkably well and ongoing refinements will generate
further insight into non-perturbative QCD mechanisms in the generation of 
hadronic final states at high energies.
\begin{table*}[htb]
  \vspace*{-0.75cm}
  \caption{H1 preliminary results of two-parameter fits of $\anmi$ and $\asmz$
    to mean event shape variables
    using the NLO power correction approach described in the text.
    The $\chi^2$ per degree of freedom, $\chi^2/n$, and the correlation 
coefficient, $\kappa$, are also given.}
  \label{klaus}
  \begin{tabular*}{\textwidth}{c@{\extracolsep\fill}ccccc}
    \hline
%    \multicolumn{6}{c}{\rbthm H1 Preliminary}\\\hline
    $\rbthm\fmean$ & $a_F$ & $\anmi$ & $\asmz$ & $\chi^2/n$ &
    $\kappa/\%$ \\\hline
    $\mean{\tau}$ & $1$ &
    $0.480 \pm 0.028~^{+0.048}_{-0.062}$ &
    $0.1174 \pm 0.0030~^{+0.0097}_{-0.0081}$ &
    $0.5$ & $-97$ \rbtnpbr\\
    $\mean{B}$ & $1/2$ &
    $0.491 \pm 0.005~^{+0.032}_{-0.036}$ &
    $0.1106 \pm 0.0012~^{+0.0060}_{-0.0057}$ &
    $0.7$ & $-58$ \rbtnpbr\\
    $\mean{\tau_C}$ & $1$ &
    $0.475 \pm 0.003~^{+0.044}_{-0.048}$ &
    $0.1284 \pm 0.0014~^{+0.0100}_{-0.0092}$ &
    $1.3$ & $+19$ \rbtnpbr\\
    $\mean{\rho}$ & $1/2$ &
    $0.561 \pm 0.004~^{+0.051}_{-0.058}$ &
    $0.1347 \pm 0.0015~^{+0.0111}_{-0.0100}$ &
    $1.2$ & $+7$ \rbtnpbr\\
    $\mean{C}$ & $3\pi/2$ &
    $0.425 \pm 0.002~^{+0.033}_{-0.039}$ &
    $0.1273 \pm 0.0009~^{+0.0104}_{-0.0093}$ &
    $0.9$ & $+63$ \rbtnpbr\\
%    $\mean{y_{fJ}}$ & $1$ &
%    $0.258 \pm 0.004$ &
%    $0.104 \pm 0.002$ &
%    $1.9$ & $-61$ \rbtnpbr\\
\hline
  \end{tabular*}
\end{table*}
\begin{figure}[htb]
\center
\includegraphics[width=0.34\textwidth]{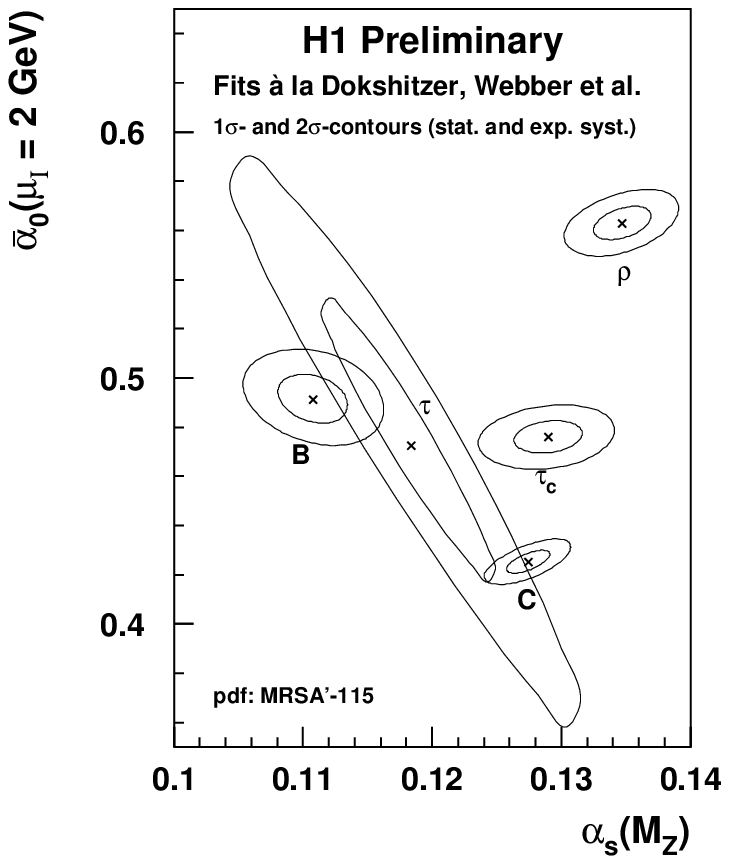}
  \hspace*{0.75cm}
\includegraphics[width=0.255\textwidth]{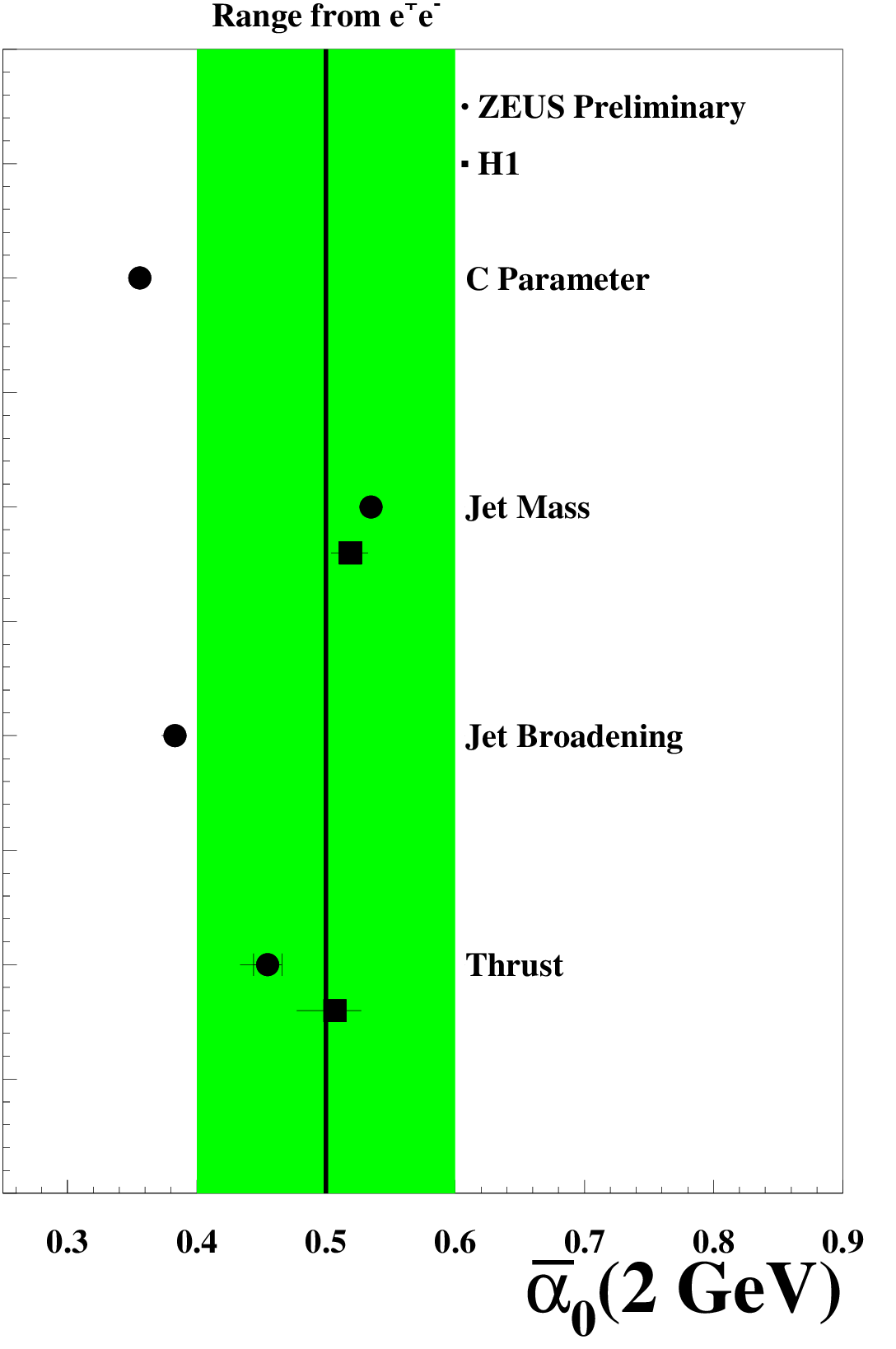}
\includegraphics[width=0.255\textwidth]{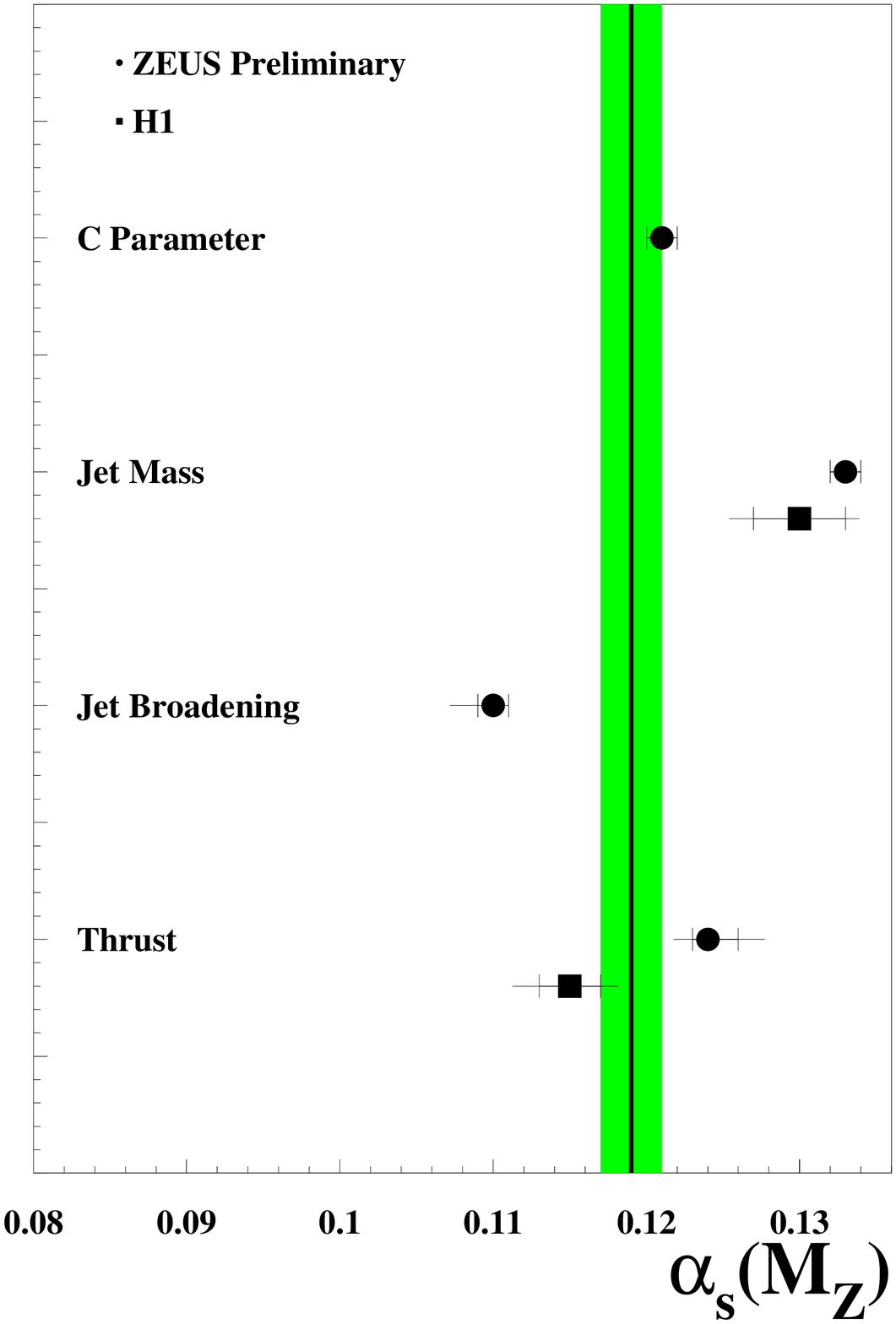}
  \vspace*{-0.75cm}
\caption{Summary of two-parameter fits of 
$\anmi$ and $\asmz$ to mean event shape variables. Left, H1 
preliminary $1\sigma$ and $2\sigma$ stat$\oplus$sys contours 
for $\tau$, $B$, $\tau_C$, $\rho$ and $C$ parameter, incorporating
the Milan factor.
Right, ZEUS preliminary fits compared to published H1 results
using leading-order power correction calculations. 
%\protect\footnote{
%(Note that the H1 published result for jet mass is 
%normalised to $Q^2$ rather than $E_{vis}^2$.)
%}
\label{rob}}
%  \vspace*{-0.75cm}
\end{figure}
%\vspace*{-0.1cm}
%\vspace*{-0.75cm}
\section{FORWARD JET PRODUCTION}
The first phase of running established HERA as the 
place to study QCD in the hitherto unexplored region of low-$x$. The rise 
of the structure function $F_2$ stimulated significant theoretical 
developments in the understanding of QCD at high energies. However, two
approaches to perturbative QCD calculations can be made, 
corresponding to whether {$\alpha_s(Q^2) \ln(Q^2)$ (DGLAP) terms or 
$\alpha_s(Q^2) \ln(1/x)$ (BFKL) terms are considered to be largest
in the perturbative splitting functions. Both approaches describe
the $F_2$ data at low-$x$: the measurement of 
forward jet production cross-sections is motivated as a test to distinguish 
between the DGLAP and BFKL approximations in low-$x$ events.

Jets with $E_T^2 \sim Q^2$ and $x_{jet}~\sgeq~x$, where 
$x_{jet}$ is the momentum fraction of the jet relative to the incoming proton,
are selected in order to enhance 
BFKL-like contributions where forward gluons may be emitted at relatively 
large $E_T$.
H1~\cite{H1jet} and ZEUS~\cite{ZEUSjet} have made measurements of 
forward jet cross-sections evaluated for $Q^2/2 < E_T^2 < 2Q^2$
as a function of $x$ for $x_{jet} > 0.035$ and $E_T > 3.5$ or 5~GeV,
with additional experimental cuts which differ in the two analyses
and account for the differences in the experimentally defined cross-sections.
There are residual ($\sim 20\%$) uncertainties in determining the 
hadron-to-parton
level corrections and therefore the measurements are presented at hadron level.
The data are shown in Figure~2 where 
the rise of $F_2$ at low-$x$ is mirrored by the rise 
of the forward jet cross-sections with decreasing $x$. 

In Figure~2(a) the data are compared to HO (Higher Order) 
BFKL calculations~\cite{martin}. 
Sub-leading corrections to the LO (Leading Order) 
cross-section are significant in the
measured region reducing the original predictions by a factor of two. 
The renormalisation scale dependence produces the largest uncertainty 
in the calculations and is indicated by the dotted/dashed
lines where the scale $\mu^2\simeq E_T^2$ is varied by a factor of four,
corresponding to a variation of the cross-section of 50\%.
The calculations describe the low-$x$ H1 and ZEUS data although there are 
hints that the low-$x$ approximations are not valid for the 
higher-$x\sim 10^{-2}$ ZEUS data. 

In Figure~2(b) the ZEUS data are compared to NLO DGLAP 
calculations~\cite{kramer}. In the upper plot, only direct contributions
of the virtual photon are included: the calculations underestimate the
cross-section by a factor of four. In the lower plot, resolved photon 
contributions, where partons from the virtual photon interact with partons
from the proton, are added: these enhance the cross-section such that 
data and theory agree. 
The renormalisation/factorisation scale dependence again produces the 
largest uncertainty and is indicated by the dashed/dotted
lines where the scale $\mu^2\simeq 2E_T^2$ is varied by a factor of three.

In conclusion, the low-$x$ forward jet data can be described either by the 
BFKL approach
{\em provided} that sub-leading terms are included or by the NLO DGLAP approach 
{\em provided} that resolved photon contributions are added. 
In both cases, the scale uncertainties are up to 50\% in the
currently measured ranges of $E_T$ and $x_{jet}$. 

\begin{figure}[htb]
\begin{minipage}[htb]{8cm}
%\vspace*{3cm}
\epsfxsize=8cm
\centering
\leavevmode
\epsfbox{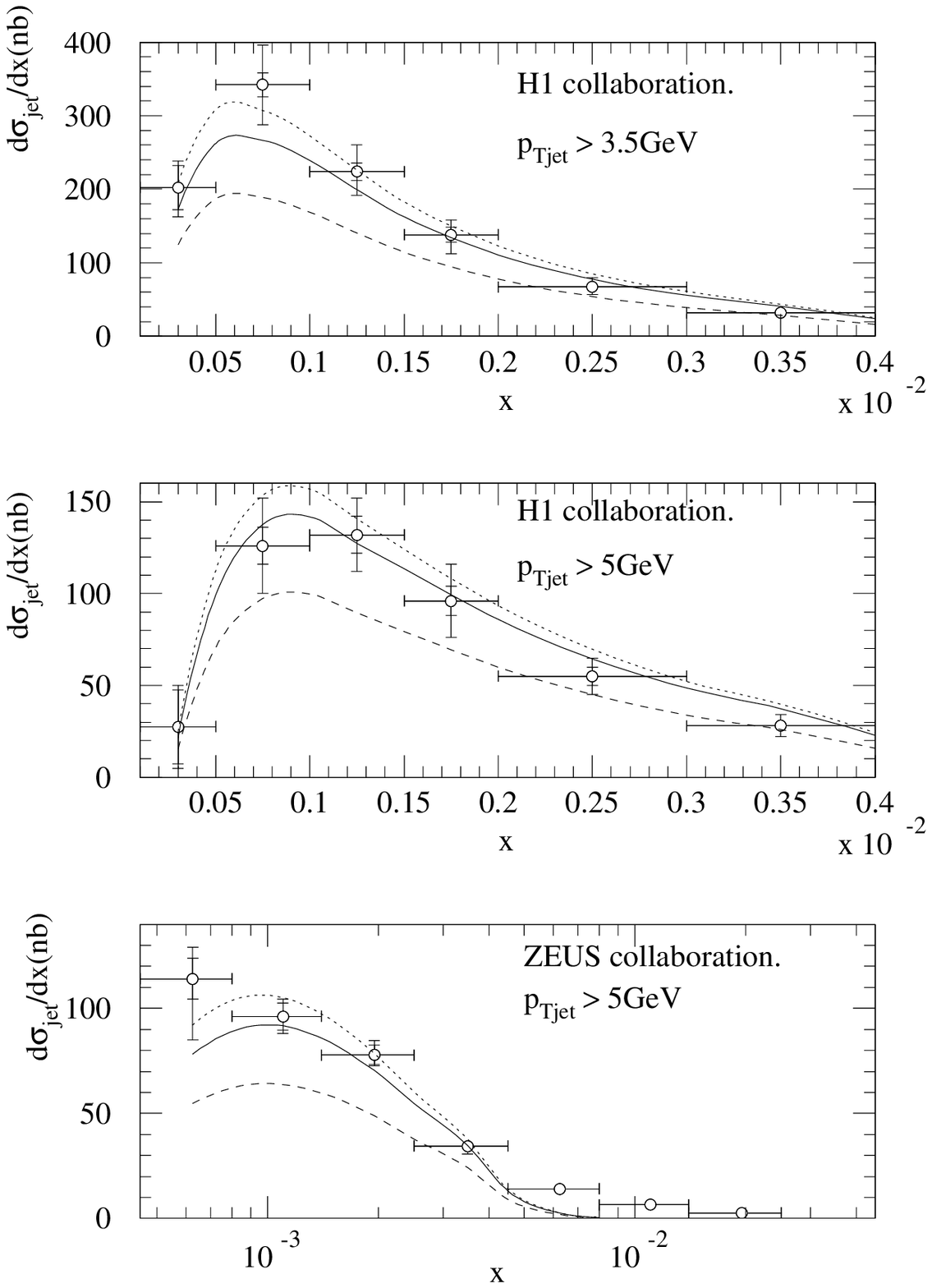}
(a) HO BFKL
\end{minipage}
\begin{minipage}[htb]{8cm}
(b) NLO DGLAP
\epsfxsize=7cm
\centering
\leavevmode
\epsfbox{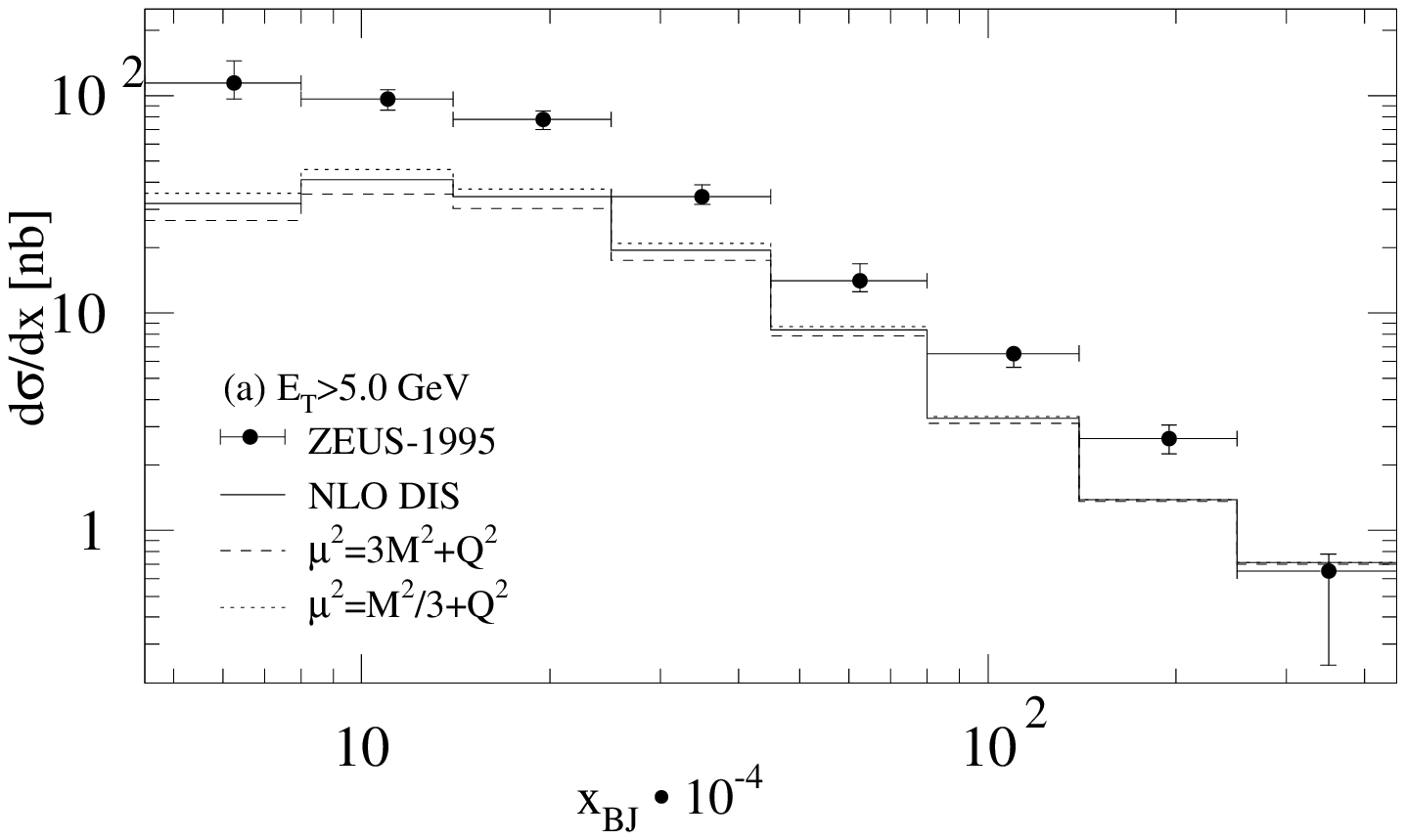}
\epsfxsize=7cm
%\centering
%\leavevmode
\epsfbox{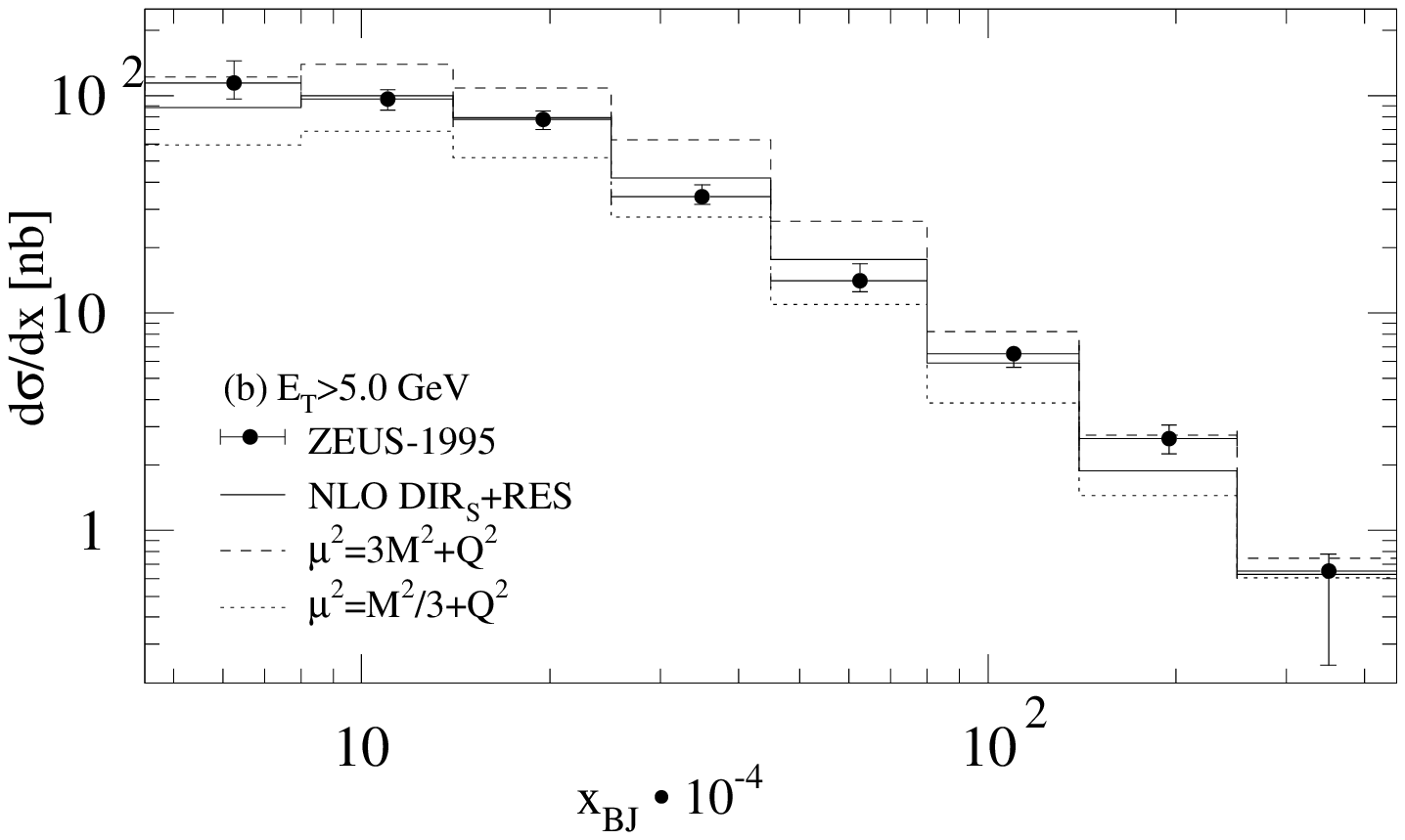}
\end{minipage}
\caption{
(a) H1 ($E_T>3.5$~GeV and $E_T>5$~GeV) and ZEUS ($E_T>5$~GeV) 
forward jet cross-sections compared to HO
BFKL calculations including sub-leading contributions~\cite{martin}  
shown on a linear scale. 
(b) ZEUS ($E_T>5$~GeV) forward jet cross-section compared to
NLO DGLAP calculations~\cite{kramer} for 
direct (above)
and direct plus resolved (below) processes shown on a logarithmic scale. 
The scale dependences, discussed in the text, are indicated by the 
dotted/dashed lines.} 
%  \vspace*{-0.75cm}
\end{figure}

\end{document}